\documentclass[conference, letterpaper]{IEEEtran}
\usepackage[switch]{lineno}
\usepackage{comment}
\IEEEoverridecommandlockouts

\usepackage{amsmath,amssymb,amsfonts}
\usepackage{algorithmic}
\usepackage[ruled,linesnumbered]{algorithm2e}
\usepackage[table,xcdraw]{xcolor}
\usepackage{graphicx}
\usepackage{textcomp}
\usepackage{xcolor}
\usepackage{nccmath}
\usepackage{environ}
\usepackage{tcolorbox}
\usepackage{hyperref}
\usepackage{soul}
\usepackage{booktabs}
\usepackage{multirow}
\usepackage{array}
\usepackage{float}
\usepackage{makecell}
\usepackage{flushend}
\usepackage{comment}
\usepackage{epstopdf} 
\SetKwFor{ForEach}{foreach}{do}{end~foreach}
\SetKwIF{If}{ElseIf}{Else}{if}{then}{else if}{else}{end~if}

\makeatletter
\newcommand{\removelatexerror}{\let\@latex@error\@gobble}
\makeatother

\begin{document}

\title{Automated Software Vulnerability Assessment with Concept Drift}

\author{\IEEEauthorblockN{Triet H. M. Le, Bushra Sabir, and
M. Ali Babar}
\IEEEauthorblockA{School of Computer Science, The University of Adelaide, Adelaide, Australia\\
Email: triet.h.le@adelaide.edu.au, bushra.sabir@adelaide.edu.au,
ali.babar@adelaide.edu.au}}

\maketitle

\begin{abstract}
Software Engineering researchers are increasingly using Natural Language Processing (NLP) techniques to automate Software Vulnerabilities (SVs) assessment using the descriptions in public repositories. However, the existing NLP-based approaches suffer from \emph{concept drift}. This problem is caused by a lack of proper treatment of new (out-of-vocabulary) terms for the evaluation of unseen SVs over time. To perform automated SVs assessment with \emph{concept drift} using SVs' descriptions, we propose a systematic approach that combines both character and word features. The proposed approach is used to predict seven Vulnerability Characteristics (VCs). The optimal model of each VC is selected using our customized time-based cross-validation method from a list of eight NLP representations and six well-known Machine Learning models. We have used the proposed approach to conduct large-scale experiments on more than 100,000 SVs in the National Vulnerability Database (NVD). The results show that our approach can effectively tackle the \emph{concept drift} issue of the SVs' descriptions reported from 2000 to 2018 in NVD even without retraining the model. In addition, our approach performs competitively compared to the existing word-only method. We also investigate how to build compact \emph{concept-drift}-aware models with much fewer features and give some recommendations on the choice of classifiers and NLP representations for SVs assessment.
\end{abstract}

\section{Introduction}
\label{sec:introduction}

Software Vulnerability (SV) is usually defined as a flaw or weakness in software code, that can potentially result in a cybersecurity attack~\cite{cve_website}. Cybersecurity attacks reportedly led to a loss of more than 50 billion dollars to the U.S. economy in 2016~\cite{cyber_crime_2016}. Different types of SVs have different levels of security threats to software-intensive systems~\cite{nayak2014some}. It is important to assess SVs for prioritizing actions so that more severe SVs are patched before exploitations~\cite{khan2018review,smyth2017software}. Automation of SVs analysis and assessment has become an important area of research efforts. Identification of SVs' characteristics is a critical task for automation. It is asserted that SVs' public repositories, such as the National Vulnerability Database (NVD), can help identify SVs characteristics by analyzing their descriptions using Natural Language Processing (NLP)~\cite{spanos2018multi,spanos2017assessment,han2017learning}.

However, the vulnerability data have the temporal property since many new terms appear in the descriptions of SVs. Such terms are a result of the release of new technologies/products or discovery of a zero-day attack or SV; for example, the NVD received more than 13,000 new SVs in 2017~\cite{nvd_website}. The appearance of new concepts makes the vulnerability data and patterns change over time~\cite{williams2018analyzing,murtaza2016mining,neuhaus2010security}, which is known as \emph{concept drift}~\cite{bullough2017predicting}. For example, the keyword Android has only started appearing in NVD since 2008, the year when Google released Android. It is being recognized that such new SVs terms cause problems for building the vulnerability assessment models.

Some previous studies~\cite{spanos2018multi,almukaynizi2019patch,bozorgi2010beyond} have suffered from \emph{concept drift} by mixing the new and old SVs in the model validation step, which can lead to biased results as such approach accidentally merges the new information with the existing one. Moreover, the previous work of SVs analysis~\cite{spanos2018multi,spanos2017assessment,almukaynizi2019patch,bozorgi2010beyond,toloudis2016associating,yamamoto2015text,edkrantz2015predicting} used predictive models with only word features without reporting how to handle the new or extended concepts (e.g., new versions of the same software) in the new SVs' descriptions. The research on machine translation~\cite{luong2014addressing,huang2011using,liu2018context,razmara2013graph} has shown that the unseen (Out-of-Vocabulary (OoV)) terms can make existing word-only models less robust to future prediction due to their missing information. For vulnerability prediction, Zhoubing~\cite{han2017learning} did use random embedding vectors to represent the OoV words, which still discards the relationship between the new and old concepts. Such observations motivated us to tackle the research problem ``\textbf{How to handle the \emph{concept drift} issue of the vulnerability descriptions in public repositories to improve the robustness of automated SVs assessment?}'' It appears to us that it is important to address the issue of \emph{concept drift} to enable practical applicability of automated vulnerability assessment tools. To the best of our knowledge, there has been no existing work to systematically address the \emph{concept drift} issue in SVs assessment.

To perform SVs assessment with \emph{concept drift} using the vulnerability descriptions in public repositories, we present a Machine Learning (ML) model that utilizes both character-level and word-level features. We also propose a customized time-based version of cross-validation method for model selection and validation. Our cross-validation method splits the data by year to embrace the temporal relationship of SVs. We evaluate the proposed model on the prediction of seven Vulnerability Characteristics (VCs), i.e., Confidentiality, Integrity, Availability, Access Vector, Access Complexity, Authentication, and Severity. Our key contributions are:

\begin{enumerate}
  \item We demonstrate the \emph{concept drift} issue of SVs using concrete examples from NVD.
  \item We investigate a customized time-based cross-validation method to select the optimal ML models for SVs assessment. Our method can help prevent future vulnerability information from being leaked into the past in model selection and validation steps.
  \item We propose and extensively evaluate an effective Character-Word Model (CWM) to assess SVs using the descriptions with \emph{concept drift}. We also investigate the performance of low-dimensional CWM models. We provide a GitHub repository\footnote{https://github.com/lhmtriet/MSR2019} containing our models and associated source code.
\end{enumerate}

\textbf{Paper structure}. Section~\ref{sec:background} introduces the vulnerability description and VCs. Section~\ref{sec:methodology} describes our proposed approach. Section~\ref{sec:expt_setup} presents the experimental design of this work. Section~\ref{sec:results} analyzes the experimental results and discusses the findings. Section~\ref{sec:discussion} identifies the threats to validity. Section~\ref{sec:related_work} covers the related works. Section~\ref{sec:conclusions} concludes and suggests some future directions.

\section{Background}
\label{sec:background}

\begin{figure}[t]
    \centering
    {%
    \setlength{\fboxsep}{0pt}%
    \setlength{\fboxrule}{0.5pt}%
    \fbox{\includegraphics[width=\linewidth,keepaspectratio]{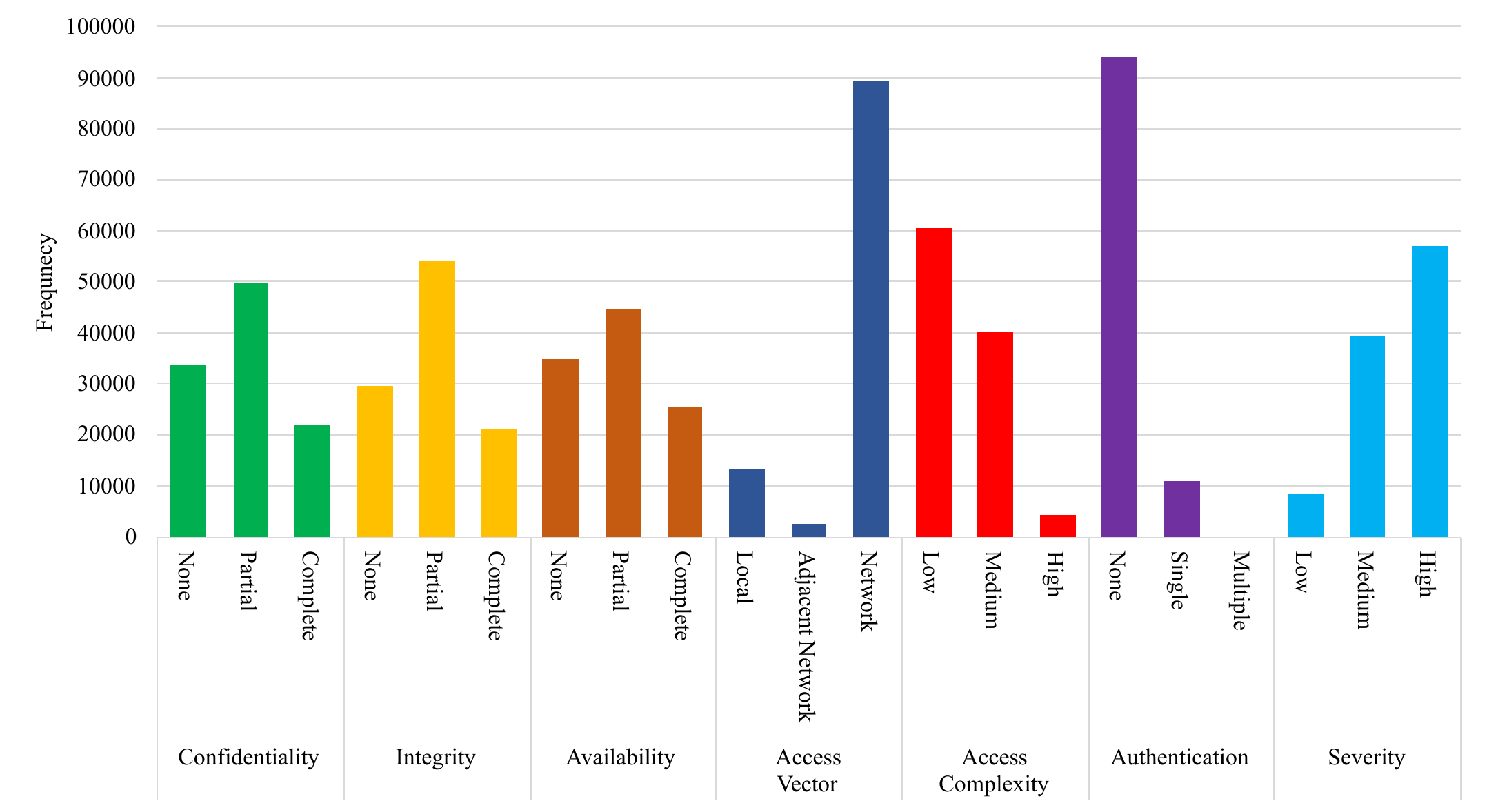}
    }}
    \caption{Frequencies of each class of the seven VCs.}
    \label{fig:vc_freqs}
\end{figure}

National Vulnerability Database~\cite{nvd_website} (NVD) is a well-known public repository that contains a huge amount of SVs information that is considered trustworthy as NVD is maintained by governmental bodies (National Cyber Security and Division of the United States Department of Homeland Security). NVD inherits the unique vulnerability identifiers and descriptions from Common Vulnerabilities and Exposures (CVE)~\cite{cve_website}. NVD also adds an evaluation to each SV using the Common Vulnerability Scoring System (CVSS)~\cite{cvss_v3,cvss_website}. Currently, there are three versions of CVSS, in which the latest version (i.e., the third version) was introduced in 2015. The second CVSS version (CVSS 2) is also maintained.

In CVSS 2, an SV is evaluated based on three main criteria: Impact, Exploitability and Severity. Impact and Exploitability refer to the threats and exploitation procedures of each SV. Severity determines the level of severity of an SV based on Impact and Exploitability. The first two criteria can be further decomposed into (Confidentiality, Integrity, Availability) and (Access Vector, Access Complexity, Authentication), respectively (cf. Fig.~\ref{fig:vc_freqs}). There are three separate values for each of the seven VCs. From the perspective of ML, this is a multi-class classification problem, which can be solved readily using ML algorithms. It is noted that Access Vector, Access Complexity and Authentication characteristics suffer the most from the issue of imbalanced data, in which the number of elements in the minority class is much smaller compared to those of the other classes.

\section{The Proposed Approach}
\label{sec:methodology}

\subsection{Approach Overview}
\label{subsec:method_overview}

\begin{figure*}[t]
    \centering
    {%
    \setlength{\fboxsep}{0pt}%
    \setlength{\fboxrule}{0.5pt}%
    \fbox{\includegraphics[width=13cm,keepaspectratio]{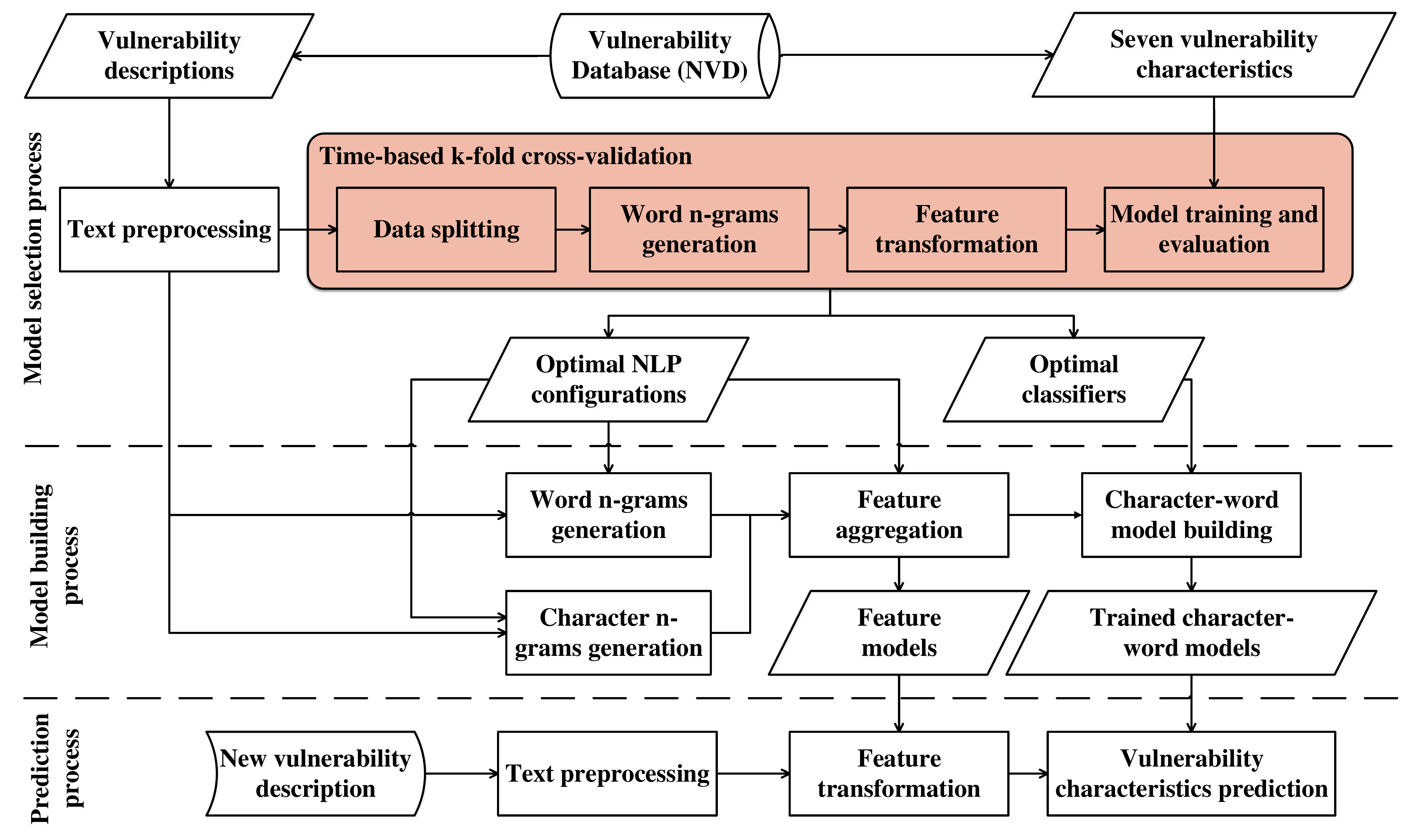}
    }}
    \caption{Main workflow of our proposed model for vulnerability assessment with \emph{concept drift}.}
    \label{fig:workflow}
\end{figure*}

The overall workflow of our proposed approach is given in Fig.~\ref{fig:workflow}. Our approach consists of three processes: model selection, model building and prediction. The first two processes work on the training set, while the prediction process performs on either a separate testing set or new vulnerability descriptions. The first model selection has two steps: Text preprocessing and Time-based k-fold cross-validation. Text preprocessing step (cf. section~\ref{subsec:preprocessing}) is necessary to reduce the noise in the text to build a better assessment model. Next, the preprocessed text enters the time-based k-fold cross-validation step to select the optimal classifier and NLP representations for each VC. It should be noted that this step only tunes the word-level models instead of the combined models of both word and character features. One reason is that the search space of the combined model is much larger than that of the word-only model since we at least have to consider different NLP representations for character-level features. The computational resource to extract character-level n-grams is also more than that of word-level counterparts. Section~\ref{subsec:time_cross-validation} provides more details about the time-based k-fold cross-validation method.

Next comes the model building process with four main steps: (\emph{i}) word n-grams generation, (\emph{ii}) character n-grams generation, (\emph{iii}) feature aggregation and (\emph{iv}) character-word model building. Steps (\emph{i}) and (\emph{ii}) use the preprocessed text in the previous process to generate word and character n-grams based on the identified optimal NLP representations of each VC. The word n-grams generation step (\emph{i}) here is the same as the one in the time-based k-fold cross-validation of the previous process. An example of the word and character n-grams in our approach is given in Table~\ref{tab:feature_examples}. Such character n-grams increase the probability of capturing parts of OoV terms due to \emph{concept drift} in vulnerability descriptions. Subsequently, both levels of the n-grams and the optimal NLP representations are input into the feature aggregation step (\emph{iii}) to extract the features from the preprocessed text using our proposed algorithm in section~\ref{subsec:feature_aggregation}. This step also combines the aggregated character and word vocabularies with the optimal NLP representations of each VC to create the feature models. We save such models to transform the data of future prediction. In the last step (\emph{iv}), the extracted features are trained with the optimal classifiers found in the model selection process to build the complete character-word models for each VC to perform automated vulnerability assessment with \emph{concept drift}.

In the prediction process, new vulnerability description is first preprocessed using the same text preprocessing step. Then, the preprocessed text is transformed to create the features by the feature models saved in the model building process. Finally, the trained character-word models use such features to determine each VC.

\begin{table}[h]
  \centering
  \caption{Word and character n-grams extracted from the sentence ``Hello World''. '\_' represents a space.}
    \begin{tabular}{|c|l|l|}
    \hline
    \textbf{n-grams} & \multicolumn{1}{c|}{\textbf{Word}} & \multicolumn{1}{c|}{\textbf{Character}} \\
    \hline
    1 & Hello, World & H, e, l, l, o, W, o, r, l, d \\
    \hline
    2 & Hello World & He, el, ll, lo, o\_, \_W, Wo, or, rl, ld \\
    \hline
    \end{tabular}%
  \label{tab:feature_examples}%
\end{table}%

\subsection{Text Preprocessing of Vulnerability Descriptions}
\label{subsec:preprocessing}

The text preprocessing is an important step for any NLP task~\cite{kao2007natural}. We use the following text preprocessing techniques: (\emph{i}) removal of stop words and punctuations, (\emph{ii}) conversion to lowercase and (\emph{iii}) stemming. The stop words are combined from the default lists of \emph{scikit-learn}~\cite{pedregosa2011scikit} and \emph{nltk}~\cite{loper2002nltk} libraries. We only remove the punctuations followed by at least one space or the ones at the end of a sentence. This punctuation removal method keeps the important words in the software and security contexts such as ``\emph{input.c}'', ``\emph{man-in-the-middle}'', ``\emph{cross-site}''.

Subsequently, the stemming step is done using the Porter Stemmer algorithm~\cite{porter1980algorithm} in \emph{nltk} library. Stemming is needed to avoid two or more words with the same meaning but in different forms (e.g., ``\emph{allow}'' vs. ``\emph{allows}''). The main goal of stemming is to retrieve consistent features (words), thus any algorithm that can return each word's root should work. Researchers may use lemmatization, which is relatively slower as it also considers the surrounding context.

\subsection{Model Selection with Time-based k-Fold Cross-Validation}
\label{subsec:time_cross-validation}

\begin{figure}[t]
    \centering
    {%
    \setlength{\fboxsep}{0pt}%
    \setlength{\fboxrule}{0.5pt}%
    \fbox{\includegraphics[width=8.3cm,keepaspectratio]{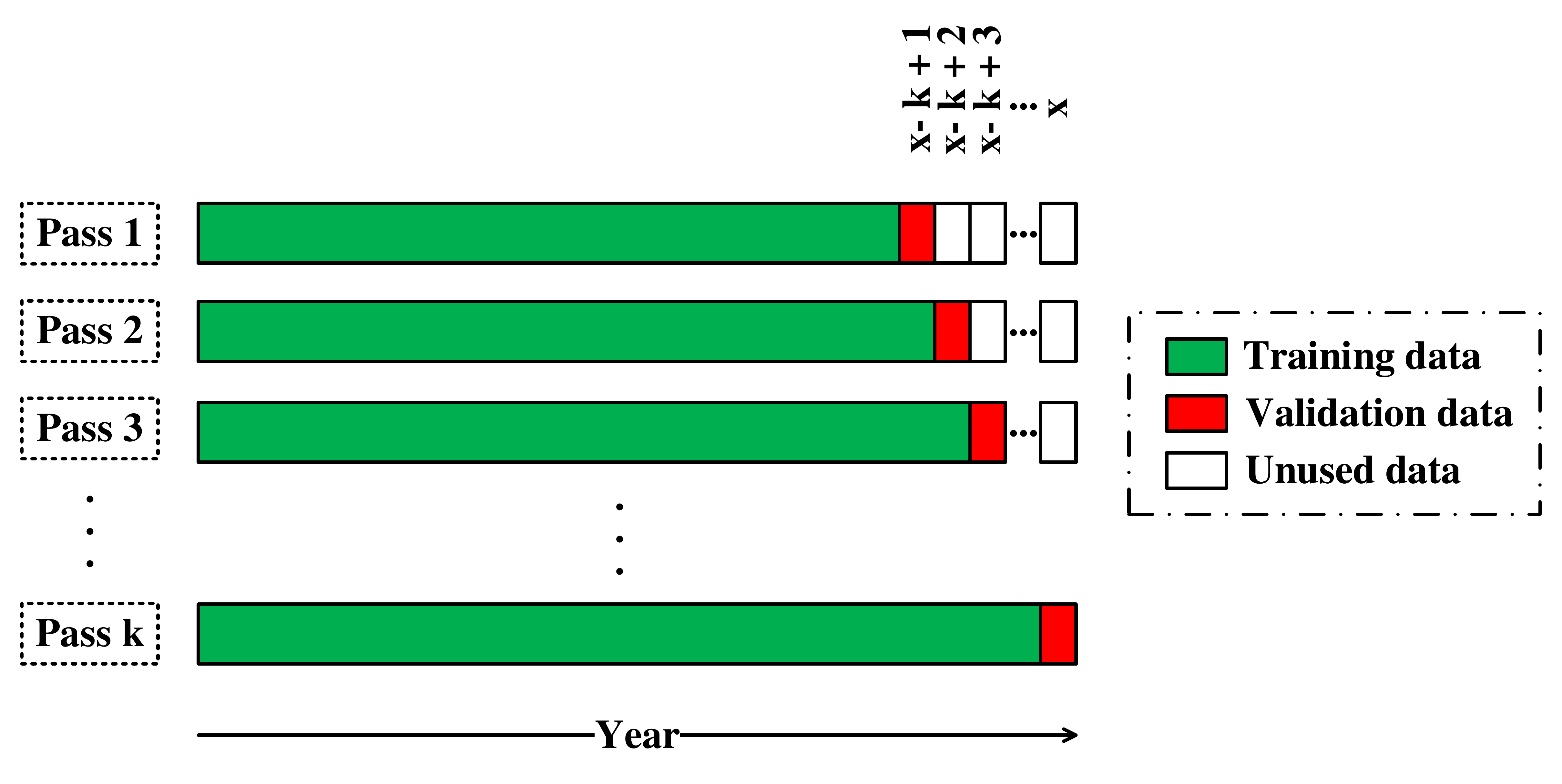}
    }}
    \caption{Our proposed time-based cross-validation method. \textbf{Note}: x is the final year in the original training set, k is the number of cross-validation folds.}
    \label{fig:time_validation}
\end{figure}

We propose a time-based cross-validation method (cf. Fig.~\ref{fig:time_validation}) to select the best classifiers and NLP representations for each VC. The idea has been inspired by the time-series domain~\cite{bergmeir2012use}. As shown in Fig.~\ref{fig:workflow}, our method has four steps: (\emph{i}) data splitting, (\emph{ii}) word n-grams generation, (\emph{iii}) feature transformation, and (\emph{iv}) model training and evaluation. Data splitting explicitly considers the time order of SVs to ensure that in each pass/fold, the new information of the validation set does not exist in the training set, which maintains the temporal property of SVs. The new terms can appear in different time during a year; thus, the preprocessed text in each fold is split by the year explicitly, not by equal sample size, e.g., SVs from 1999 to 2010 are for training and those of 2011 are for validation in a pass/fold.

\begin{table}[h]
  \centering
  \caption{Eight configurations of NLP representations used for model selection. \textbf{Note}: '\checkmark' is selected, '-' is non-selected.}
    \begin{tabular}{|c|c|c|}
    \hline
    \textbf{Configuration} & \textbf{Word n-grams} & \textbf{tf-idf} \\
    \hline
    1     & 1     & - \\
    \hline
    2     & 1     & \checkmark \\
    \hline
    3     & 1-2   & - \\
    \hline
    4     & 1-3   & - \\
    \hline
    5     & 1-4   & - \\
    \hline
    6     & 1-2   & \checkmark \\
    \hline
    7     & 1-3   & \checkmark \\
    \hline
    8     & 1-4   & \checkmark \\
    \hline
    \end{tabular}%
  \label{tab:model_configurations}%
\end{table}%

After data splitting in each fold, we use the training set to generate the word n-grams. Subsequently, with each of the eight NLP configurations in Table~\ref{tab:model_configurations}, the feature transformation step uses the word n-grams as the vocabulary to transform the preprocessed text of both training and validation sets into the features for building a model. We create the NLP configurations from various values of n-grams combined with either term frequency or tf-idf measure. Uni-gram with term frequency is also called Bag-of-Words (BoW). These NLP representations have been selected since they are popular and have performed well for SVs analysis~\cite{spanos2018multi,almukaynizi2019patch,yamamoto2015text}. For each NLP configuration, the model training and evaluation step trains six classifiers (cf. section~\ref{subsec:ml_models}) on the training set and then evaluates the models on the validation set using different evaluation metrics (cf. section~\ref{subsec:evaluation_metrics}). The model with the highest average cross-validated performance is selected for the current VC. The process is repeated for every VC, then the optimal classifiers and NLP representations are returned for all seven VCs.

\subsection{Feature Aggregation Algorithm}
\label{subsec:feature_aggregation}

We propose Algorithm~\ref{algo:feature_aggregation} to combine word and character n-grams in the model building process to create the features for our character-word model. Six inputs of the algorithm are (\emph{i}) input descriptions, (\emph{ii}) word n-grams, (\emph{iii}) character n-grams, (\emph{iv}) the minimum, (\emph{v}) the maximum number of character n-grams, and (\emph{vi}) the optimal NLP configuration of the current VC. The main output is a feature matrix containing the term weights of the documents transformed by the aggregated character and word vocabularies to build the character-word models. We also output the character and word feature models for future prediction.

\begin{algorithm}[htbp]
    \caption{Feature aggregation algorithm to transform the documents with the aggregated word and character-level features.}
    \label{algo:feature_aggregation}
    \DontPrintSemicolon

    \KwIn{List of vulnerability descriptions: ${D}_{in}$\\
    Set of word-level n-grams: ${F}_{w}=\{{{f}_{1w}},{{f}_{2w}},...,{{f}_{nw}}\}$\\
    Set of character-level n-grams: ${F}_{c}=\{{{f}_{1c}},{{f}_{2c}},...,{{f}_{mc}}\}$\\
    The minimum and maximum character n-grams: ${min}_{n-gram}$ and ${max}_{n-gram}$\\
    The optimal NLP configuration of the current VC: \emph{config}}
    \KwOut{The aggregated data matrix: ${\mathbf{X}}_{agg}$\\
    The word and character feature models: ${\text{model}}_{w}, {\text{model}}_{c}$}

    $slt\_chars \leftarrow \emptyset$\;

    \ForEach{${f}_{i}\in {F}_{c}$}{
        $tokens \leftarrow {f}_{i}$ trimmed and split by space\;

        \If{(size of $tokens$ = 1) {\bf and} ((length of the first element in $tokens$) $>$ 1)}{
            $slt\_chars \leftarrow slt\_chars + \{tokens\}$\;
        }
    }

    $diff\_words \leftarrow {F}_{w} - slt\_chars$\;

    ${\text{model}}_{w} \leftarrow \text{Feature\_transformation}(diff\_words, config)$\;

    ${\text{model}}_{c} \leftarrow \text{Feature\_transformation}(slt\_chars, {min}_{n-gram} - 1, {max}_{n-gram}, config)$\;

    ${\mathbf{X}}_{word} \leftarrow {D}_{in}$ transformed with ${\text{model}}_{w}$\;
    ${\mathbf{X}}_{char} \leftarrow {D}_{in}$ transformed with ${\text{model}}_{c}$\;
    ${\mathbf{X}}_{agg} \leftarrow \text{horizontal\_append}({\mathbf{X}}_{word}, {\mathbf{X}}_{char})$\;

    \Return ${\mathbf{X}}_{agg}$, ${\text{model}}_{w}$, ${\text{model}}_{c}$

\end{algorithm}

Steps 2-7 of the algorithm filter the character features. More specifically, step 3 removes (trims) the spaces from both ends of each feature. Then, we split such feature by space(s) to determine how many words to which the character(s) belongs. Subsequently, steps 4-6 retain only the character features that are parts of single words (size of $tokens$ = 1), except the single characters such as $x$, $y$, $z$ ((length of the first element in $tokens$) $>$ 1). The n-gram characters with space(s) in between represent more than one word, which can make the classifier more prone to overfitting. Similarly, single characters are too short and they can belong to too many words, which is likely to make the model hardly generalizable. In fact, '\emph{a}' is a meaningful single character, but it has been already removed as a stop word. The characters can even represent a whole word (e.g., ``\emph{attack}'' token with ${max}_{n-gram}\ge \,6$). In such cases, step 8 removes the duplicated word-level features (${F}_{w} - slt\_chars$). Based on the assumption that unseen or misspelled terms can share common characters with existing words, such choice can enhance the probability of the model capturing the OoV words in the new descriptions. Retaining only the character features also helps reduce the number of features and the model overfitting. After that, steps 9-10 define the feature models ${\text{model}}_{w}$ and ${\text{model}}_{c}$ using the word ($diff\_words$) and character ($slt\_chars$) vocabularies, respectively, along with the NLP configurations to transform the input documents into feature matrices for building the model. Steps 11-12 then use the two defined word and character models to actually transform the input documents   into feature matrices ${\mathbf{X}}_{word}$ and ${\mathbf{X}}_{word}$, respectively. Step 13 concatenates two feature matrices by columns. Finally, step 14 returns the final aggregated feature matrix ${\mathbf{X}}_{agg}$ along with both word and character feature models namely ${\text{model}}_{w}$ and ${\text{model}}_{c}$.

\section{EXPERIMENTAL DESIGN}
\label{sec:expt_setup}

All the classifiers and NLP representations (n-grams, term frequency, tf-idf) in this work have been implemented in \emph{scikit-learn}~\cite{pedregosa2011scikit} and \emph{nltk}~\cite{loper2002nltk} libraries in Python. Our code ran on a fourth-generation Intel Core i7-4200HQ CPU (four cores) running at 2.6 GHz with 16 GB of RAM.

\subsection{Research Questions}
\label{subsec:rqs}

Our research aims at addressing the \emph{concept drift} issue in SVs' descriptions to improve the robustness of both model selection and prediction steps. We have evaluated our two-phase character-word model. The first phase selects the optimal models for each VC. The second phase incorporates the character features to build character-word models. We raise and answer four Research Questions (RQs):

\begin{itemize}
  \item \textbf{RQ1}: \emph{Is our time-based cross-validation more effective than a non-temporal method to handle concept drift in the model selection step for vulnerability assessment?} To answer RQ1, we first identify the new terms in the vulnerability descriptions. We associate such terms with their release or discovery years. We then use qualitative examples to demonstrate the information leakage in the non-temporal model selection step. We also quantitatively compare the effectiveness of the proposed time-based cross-validation method with a traditional non-temporal one to address the temporal relationship in the context of vulnerability assessment.
  \item \textbf{RQ2}: \emph{Which are the optimal models for multi-classification of each vulnerability characteristic?} To answer RQ2, we present the optimal models (i.e., classifiers and NLP representations) using word features for each VC selected by a five-fold time-based cross-validation method (cf. section~\ref{subsec:time_cross-validation}). We also compare the performance of different classes of models (single vs. ensemble) and NLP representations to give recommendations for future use.
  \item \textbf{RQ3}: \emph{How effective is our character-word model to perform automated vulnerability assessment with concept drift?} For RQ3, we first demonstrate how the OoV phrases identified in RQ1 can affect the performance of the existing word-only models. We then highlight the ability of the character features to handle the \emph{concept drift} issue of SVs. We also compare the performance of our character-word model with those of the word-only (without handling \emph{concept drift}) and character-only models.
  \item \textbf{RQ4}: \emph{To what extent can low-dimensional model retain the original performance?} The features of our proposed model in RQ3 are high-dimensional and sparse. Hence, we evaluate the dimensionality reduction technique (i.e., Latent Semantic Analysis~\cite{deerwester1990indexing}) and the sub-word embeddings (i.e., fastText~\cite{bojanowski2017enriching,joulin2016bag}) to show how much information of the original model is approximated in lower dimensions. The work done for answering RQ4 facilitates the building of more efficient \emph{concept-drift}-aware predictive models.
\end{itemize}

\subsection{Dataset}
\label{subsec:dataset}

We retrieved 113,292 SVs from NVD in JSON format. The dataset contains the SVs from 1988 to 2018. We discarded 5926 SVs that contain ``** REJECT **'' in their descriptions since they have been confirmed duplicated or incorrect by experts. Seven VCs of CVSS 2 (cf. section~\ref{sec:background}) were used as our SVs assessment metrics. It turned out that there are 2242 SVs without any value of CVSS 2. Therefore, we also removed such SVs from our dataset. Finally, we obtained a dataset containing 105,124 SVs along with their descriptions and the values of seven VCs indicated previously. For evaluation purposes, we followed the work in~\cite{spanos2018multi} to use the year of 2016 to divide our dataset into training and testing sets with the sizes of 76,241 and 28,883, respectively. The primary reason for splitting the dataset based on the time order is to consider the temporal relationship of SVs.

\subsection{Vulnerability Classification Machine Learning Models}
\label{subsec:ml_models}

To solve our multi-class classification problem, we used six well-known ML models. These classifiers have achieved great results in recent data science competitions such as Kaggle~\cite{kaggle_website}. We provide brief descriptions and the hyperparameters of each classifier below.
\begin{itemize}
  \item Na\"ive Bayes (NB)~\cite{russell2002artificial} is a simple probabilistic model that is based on Bayes' theorem. This model assumes that all the features are conditionally independent with respect to each other. In this study, NB has no tuning hyperparameter during the validation step.
  \item Logistic Regression (LR)~\cite{walker1967estimation} is a linear classifier in which the logistic function is used to convert the linear output into probability. The one-vs-rest scheme is applied to split the multi-class problem into multiple binary classification problems. In this work, we select the optimal value of the regularization parameter for LR from the list of values: 0.01, 0.1, 1, 10, 100.
  \item Support Vector Machine (SVM)~\cite{cortes1995support} is a classification model in which a maximum margin is determined to separate the classes. For NLP, the linear kernel is preferred because of its more scalable computation and sparsity handling~\cite{basu2003support}. The tuning regularization values of SVM are the same as LR.
  \item Random Forest (RF)~\cite{ho1995random} is a bagging model in which multiple decision trees are combined to reduce the variance and sensitivity to noise. The complexity of RF is mainly controlled by (\emph{i}) the number of trees, (\emph{ii}) maximum depth, and (\emph{iii}) maximum number of leaves. (\emph{i}) tuning values are: 100, 300, 500. We set (\emph{ii}) to \emph{unlimited}, which makes the model the highest degree of flexibility and easier to adapt to new data. For (\emph{iii}), the tuning values are 100, 200, 300 and \emph{unlimited}.
  \item XGBoost - Extreme Gradient Boosting (XGB)~\cite{chen2016xgboost} is a variant Gradient Boosting Tree Model (GBTM) in which multiple weak tree-based classifiers are combined and regularized to enhance the robustness of the overall model. Three hyperparameters of XGB that require tuning are the same as RF. It should be noted that the \emph{unlimited} value of the maximum number of leaves is not applicable to XGB.
  \item Light Gradient Boosting Machine (LGBM)~\cite{ke2017lightgbm} is a light-weight version of GBTM. Its main advantage is the scalability since the sub-trees are grown in a leaf-wise manner rather than depth-wise of other GBT algorithms. Three hyperparameters of LGBM that require tuning are the same as XGB.
\end{itemize}

In this work, we consider NB, LR and SVM as single models, while RF, XGB and LGBM as ensemble models.

\subsection{Evaluation Metrics}
\label{subsec:evaluation_metrics}

Our multi-class classification problem can be decomposed into multiple binary classification problems. To define the standard evaluation metrics for a binary problem~\cite{spanos2018multi,spanos2017assessment,han2017learning}, we first describe four possibilities as follows.
\begin{itemize}
  \item True positive (\emph{TP}): The classifier correctly predicts that an SV has a particular characteristic.
  \item False positive (\emph{FP}): The classifier incorrectly predicts that an SV has a particular characteristic.
  \item True negative (\emph{TN}): The classifier correctly predicts that an SV does not have a particular characteristic.
  \item False negative (\emph{FN}): The classifier incorrectly predicts that an SV does not have a particular characteristic.
\end{itemize}

Based on \emph{TP}, \emph{FP}, \emph{TN}, \emph{FN}, \emph{Accuracy}, \emph{Precision}, \emph{Recall} and \emph{F-Score} can be defined accordingly in~\eqref{eq:accuracy},~\eqref{eq:precision},~\eqref{eq:recall},~\eqref{eq:f_score}.

\begin{equation}\label{eq:accuracy}
  Accuracy\,\,=\,\,\frac{TP\,+\,TN}{TP+\,FP\,+\,TN\,+\,FN}
\end{equation}

\begin{equation}\label{eq:precision}
  Precision\,=\,\frac{TP}{TP\,+\,FP}
\end{equation}

\begin{equation}\label{eq:recall}
  Recall\,=\,\frac{TP}{TP\,+\,FN}
\end{equation}

\begin{equation}\label{eq:f_score}
  F-Score\,=\,\frac{2\,\times \,Precision\,\times \,Recall}{Precision\,+\,Recall}
\end{equation}

Whilst \emph{Accuracy} measures the global performance of all classes, \emph{F-Score} (a harmonic mean of \emph{Precision} and \emph{Recall}) evaluates each class separately. Such local estimate like F-Score is more favorable to the imbalanced VCs such as Access Vector, Access Complexity, Authentication, Severity (cf. Fig.~\ref{fig:vc_freqs}). In fact, there are several variants of \emph{F-Score} for multi-class classification problem namely \emph{Micro}, \emph{Macro} and Weighted\emph{ F-Scores}. In the case of multi-class classification, \emph{Micro F-Score} is actually the same as \emph{Accuracy}. For \emph{Macro} and \emph{Weighted F-Scores}, the former does not consider class distribution (the number of elements in each class) for computing \emph{F-Score} of each class; whereas, the latter does. To account for the balanced and imbalanced VCs globally and locally, we use \emph{Accuracy}, \emph{Macro}, and \emph{Weighted F-Scores} to evaluate our models. For model selection, if there is a performance tie among models regarding \emph{Accuracy} and/or \emph{Macro F-Score}, \emph{Weighted F-Score} is chosen as the discriminant criterion. The reason is that \emph{Weighted F-Score} can be considered a compromise between \emph{Macro F-Score} and \emph{Accuracy}. If the tie still exists, the less complex model with the smaller number of hyperparameters is selected as per the Occam's razor principle~\cite{blumer1987occam}. In the last tie scenario, the model with shorter training time is chosen.

\section{Experimental Results and Discussion}
\label{sec:results}

\subsection{\textbf{RQ1}: Is our time-based cross-validation more effective than a non-temporal method to handle concept drift in the model selection step for vulnerability assessment?}
\label{subsec:rq1_results}

\begin{figure}[t]
    \centering
    {%
    \setlength{\fboxsep}{0pt}%
    \setlength{\fboxrule}{0.5pt}%
    \fbox{\includegraphics[width=8.3cm,keepaspectratio]{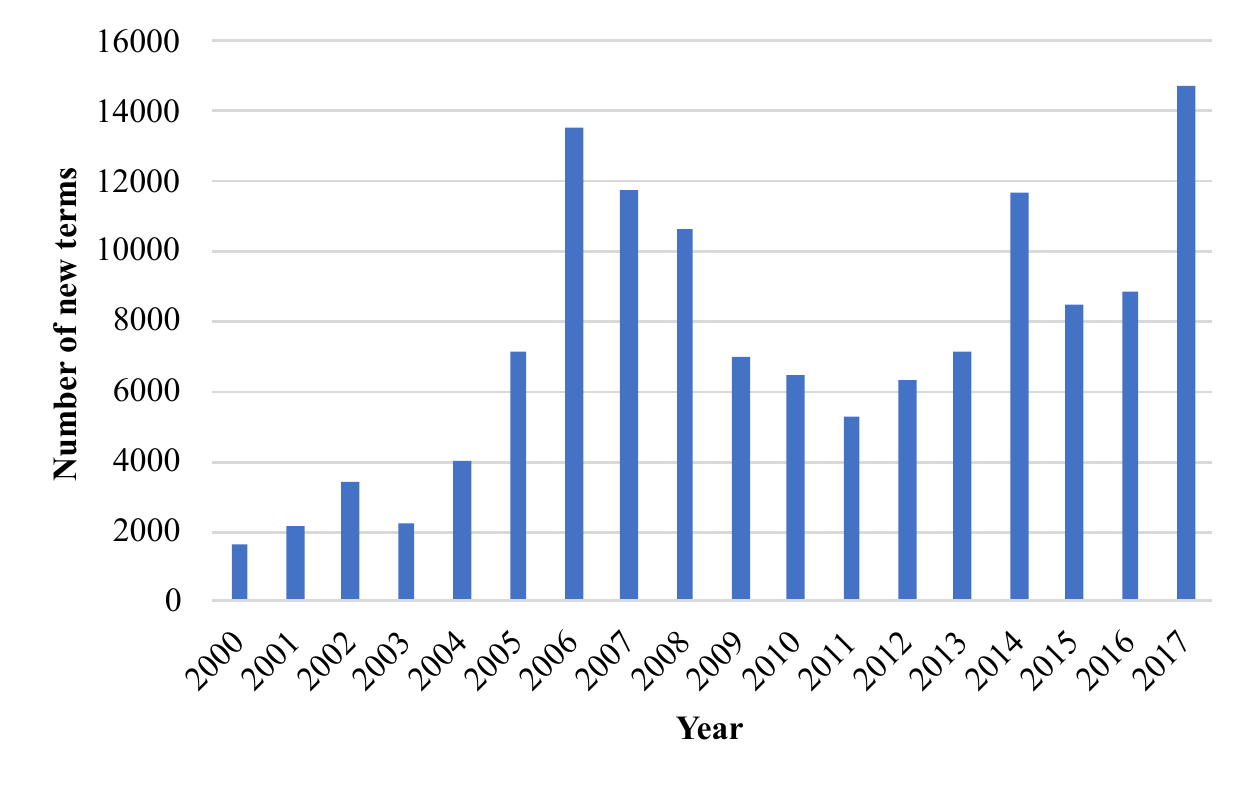}
    }}
    \caption{The number of new terms from 2000 to 2017 of vulnerability description in NVD.}
    \label{fig:new_terms}
\end{figure}

\begin{figure}[b]
    \centering
    {%
    \setlength{\fboxsep}{0pt}%
    \setlength{\fboxrule}{0.5pt}%
    \fbox{\includegraphics[width=8.3cm,keepaspectratio]{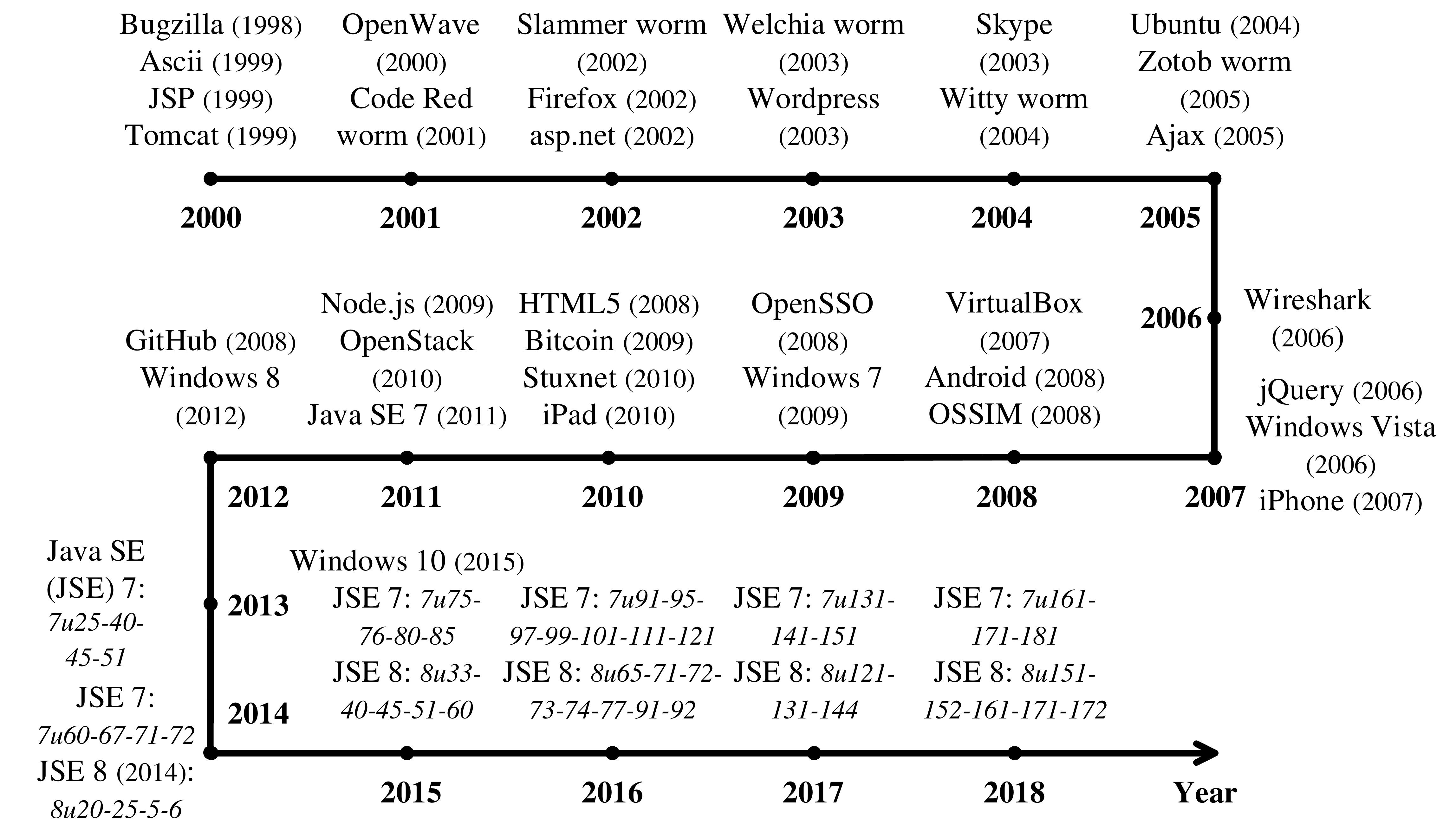}
    }}
    \caption{Examples of new terms in NVD corresponding to new products, software, cyber-attacks from 2000 to 2018. \textbf{Note}: The year of release/discovery is put in parentheses.}
    \label{fig:sv_timeline}
\end{figure}

We performed both qualitative and quantitative analyses to demonstrate the relationship between \emph{concept drift} and the model selection step of vulnerability assessment. Firstly, it is intuitive that the data of SVs intrinsically changes over time because of new products, software and attack vectors. The number of new terms appearing in the NVD description each year during the period from 2000 to 2017 is given in Fig.~\ref{fig:new_terms}. On average each year, there are 7345 new terms added to the vocabulary. Moreover, from 2015 to 2017, the number of new terms has been consistently increasing and achieved an all-time high value of 14684 in 2017. We also highlight some concrete examples about the terms appearing in the database after a particular technology, product or attack was released in Fig.~\ref{fig:sv_timeline}. There seems to be a strong correlation between the time of appearance of some new terms in the descriptions and their years of release or discovery. Such unseen terms contain many concepts about new products (e.g., \emph{Firefox}, \emph{Skype}, and \emph{iPhone}), operating systems (e.g., \emph{Android}, \emph{Windows Vista/7/8/10}), technologies (e.g., \emph{Ajax}, \emph{jQuery}, and \emph{Node.js}), attacks (e.g., \emph{Code Red}, \emph{Slammer}, and \emph{Stuxnet} worms). There are also the extended forms of existing ones such as the updated versions of Java Standard Edition (Java SE) each year. These qualitative results depict that if the time property of SVs is not considered in the model selection step, then the future terms can be mixed with past ones. Such information leakage can result in the discrepancy in the real model performance.

\begin{figure}[t]
    \centering
    {%
    \setlength{\fboxsep}{0pt}%
    \setlength{\fboxrule}{0.5pt}%
    \fbox{\includegraphics[width=8.3cm,keepaspectratio]{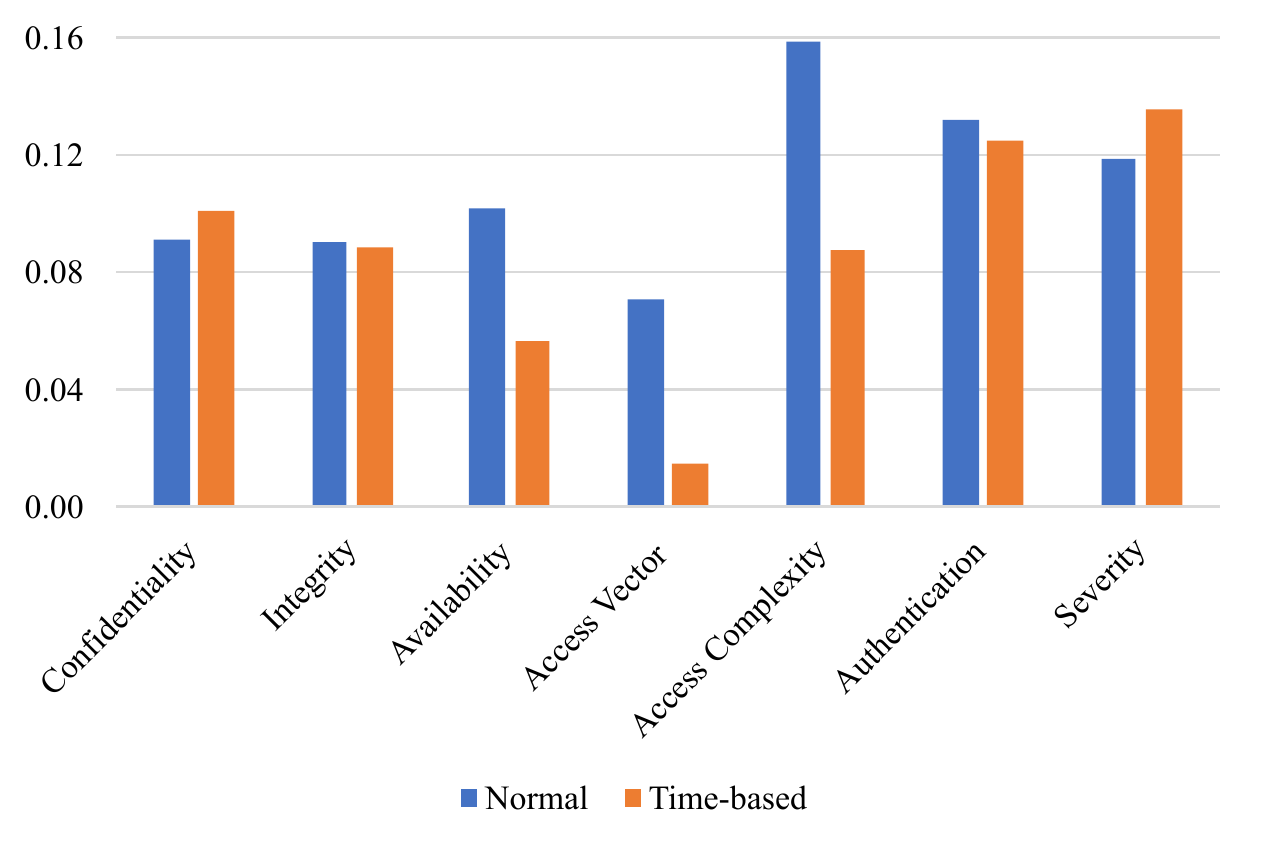}
    }}
    \caption{Difference between the validated and testing \emph{Weighted F-Scores} of our time-based and a normal cross-validation methods.}
    \label{fig:time-norm_overfitting}
\end{figure}

In fact, the main goal of the validation step is to select the optimal models that can exhibit the similar behavior on unseen data. Next, our approach quantitatively compared the degree of model overfitting using our time-based cross-validation method with a stratified non-temporal one used in~\cite{spanos2018multi,spanos2017assessment}. For each method, we computed the \emph{Weighted F-Scores} difference between the cross-validated and testing results of the optimal models found in the validation step (cf. Fig.~\ref{fig:time-norm_overfitting}). The model selection and selection criteria procedures of the normal cross-validation method are the same as our temporal one. Fig.~\ref{fig:time-norm_overfitting} shows that traditional non-temporal cross-validation was overfitted in four out of seven cases (i.e., Availability, Access Vector, Access Complexity, and Authentication). Especially, the degrees of overfitting of non-temporal validation method were 1.8, 4.7 and 1.8 times higher than those of the time-based version for Availability, Access Vector, and Access Complexity, respectively. For the other three VCs, both methods were similar, in which the differences were within 0.02. Moreover, on average, the \emph{Weighted F-Scores} on the testing set of the non-temporal cross-validation method were only 0.002 higher than our approach. This value is negligible compared to the difference of 0.02 (ten times more) in the validation step. It is noted that similar comparison also holds for non-stratified non-temporal cross-validation. Overall, both qualitative and quantitative findings suggest that the time-based cross-validation method should be preferred to lower the performance overestimation and mis-selection of predictive models due to the effect of \emph{concept drift} in the model selection step of SVs.

\begin{tcolorbox}
\textbf{The summary answer to RQ1}: The qualitative results show that many new terms are regularly added to NVD, after the release or discovery of the corresponding software products or cyber-attacks. Normal methods mixing these new terms can inflate the cross-validated model performance. Quantitatively, the optimal models found by our time-based cross-validation are also less overfitted, especially two to five times for Availability, Access Vector and Access Complexity. It is recommended that the time-based cross-validation should be adopted in the model selection step for vulnerability assessment.
\end{tcolorbox}

\subsection{RQ2: Which are the optimal models for multi-classification of each vulnerability characteristic?}
\label{subsec:rq2_results}

The answer to RQ1 has shown that the temporal cross-validation should be used for selecting the optimal models in the context of vulnerability assessment. The work to answer RQ2 presents the detailed results of the first phase of our model. To be more specific, we have used our five-fold time-based cross-validation to select the optimal word-only model for each of the seven VCs from six classifiers (cf. section IV.C) and eight NLP representations (cf. Table~\ref{tab:model_configurations}). More specifically, we have followed the guidelines of the previous work~\cite{spanos2018multi} to extract only the words appearing in more than 0.1\% of all descriptions as features for RQ2.

\begin{table}[t]
  \centering
  \caption{Optimal hyperparameters found for each classifier.}
    \begin{tabular}{|l|l|}
    \hline
    \textbf{Classifier} & \multicolumn{1}{c|}{\textbf{Hyperparameters}} \\
    \hline
    \textbf{NB} & None \\
    \hline
    \multirow{3}{*}{\textbf{LR}} & Regularization value: \\
     & + 0.1 for term frequency \\
     & + 10 for tf-idf \\
    \hline
    \multirow{2}{*}{\textbf{SVM}} & Kernel: linear \\
     & Regularization value: 0.1 \\
    \hline
    \multirow{3}{*}{\textbf{RF}} & Number of trees: 100 \\
     & Max. depth: unlimited \\
     & Max. number of leaf nodes: unlimited \\
    \hline
    \multirow{3}{*}{\textbf{XGB}} & Number of trees: 100 \\
     & Max. depth: unlimited \\
     & Max. number of leaf nodes: 100 \\
    \hline
    \multirow{3}{*}{\textbf{LGBM}} & Number of trees: 100 \\
     & Max. depth: unlimited \\
     & Max. number of leaf nodes: 100 \\
    \hline
    \end{tabular}%
  \label{tab:optimal_hyperparameters}%
\end{table}%

Firstly, each classifier was tuned using random VCs to select its optimal set of hyperparameters. Such selected hyperparameters are reported in Table~\ref{tab:optimal_hyperparameters}. It is worth noting that we have utilized local optimization as a filter to reduce the search space. We found that 0.1 was a consistently good value of regularization coefficient for SVM. Unlike SVM, for LR, 0.1 was suitable for term frequency representation; whereas, 10 performed better for the case of tf-idf. One possible explanation is that LR provides a decision boundary that is more sensitive to hyperparameter. Additionally, although tf-idf with l2-normalization helps model converge faster, it usually requires more regularization to avoid overfitting~\cite{le2018identification}. For ensemble models, more hyperparameters need tuning as mentioned in section~\ref{subsec:ml_models}. Regarding the maximum number of leaves, the optimal value for RF was \emph{unlimited}, which is expected since it would give more flexibility to the model.

\begin{table}[t]
  \centering
  \caption{Optimal models and results after the validation step. \textbf{Note}: The NLP configuration number is put in parentheses.}
    \begin{tabular}{|l|l|c|c|c|}
    \hline
    \makecell{\textbf{Vulnerability}\\ \textbf{characteristic}} & \multicolumn{1}{c|}{\makecell{\textbf{Classifier}\\ \textbf{(config)}}} & \textbf{\emph{Accuracy}} & \makecell{\textbf{\emph{Macro}}\\ \textbf{\emph{F-Score}}} & \makecell{\textbf{\emph{Weighted}}\\ \textbf{\emph{F-Score}}} \\
    \hline
    \textbf{Confidentiality} & LGBM (1) & 0.839 & 0.831 & 0.84 \\
    \hline
    \textbf{Integrity} & XGB (4) & 0.861 & 0.853 & 0.861 \\
    \hline
    \textbf{Availability} & LGBM (1) & 0.785 & 0.783 & 0.782 \\
    \hline
    \textbf{Access Vector} & XGB (7) & 0.936 & 0.643 & 0.914 \\
    \hline
    \textbf{Access Complexity} & LGBM (1) & 0.771 & 0.553 & 0.758 \\
    \hline
    \textbf{Authentication} & LR (3) & 0.973 & 0.626 & 0.972 \\
    \hline
    \textbf{Severity} & LGBM (5) & 0.814 & 0.763 & 0.811 \\
    \hline
    \end{tabular}%
  \label{tab:optimal_models}%
\end{table}%

However, for XGB and LGBM, the \emph{unlimited} value was not available. In fact, the higher value did not improve the performance, but significantly increased the computational time. As a result, we chose 100 to be the number of leaves for XGB and LGBM. Similarly, we obtained 100 as a good value for the number of trees of each ensemble model. We noticed that the maximum depth of ensemble methods was the hyperparameter that affected the validation result the most; the others did not change the performance dramatically. Finally, we got a search space of size of 336 in the cross-validation step ((six classifiers) $\times$ (eight NLP configurations) $\times$ (seven characteristics)). After using our five-fold time-based cross-validation method in section~\ref{subsec:time_cross-validation}, the optimal validation results are given in Table~\ref{tab:optimal_models}.

\begin{table}[t]
  \centering
  \caption{Average cross-validated \emph{Weighted F-scores} of term frequency vs. tf-idf grouped by six classifiers.}
    \begin{tabular}{|l|c|c|c|c|c|c|}
    \cline{2-7}
    \multicolumn{1}{l|}{}      & \multicolumn{6}{c|}{\textbf{Classifier}} \\
    \cline{2-7}
    \multicolumn{1}{l|}{} & \textbf{NB} & \textbf{LR} & \textbf{SVM} & \textbf{RF} & \textbf{XGB} & \textbf{LGBM} \\
    \hline
    \makecell[l]{\textbf{Term}\\ \textbf{frequency}} & 0.781 & \textbf{0.833} & \textbf{0.835} & \textbf{0.843} & \textbf{0.846} & \textbf{0.846} \\
    \hline
    \textbf{tf-idf} & \textbf{0.786} & 0.832 & 0.831 & 0.836 & 0.843 & 0.844 \\
    \hline
    \end{tabular}%
  \label{tab:feature_clf}%
\end{table}%

\begin{table}[b]
  \centering
  \caption{Average cross-validated \emph{Weighted F-scores} of uni-gram vs. n-grams (2 $\leq$ n $\leq$ 4) grouped by six classifiers.}
    \begin{tabular}{|l|c|c|c|c|}
    \hline
    \multicolumn{1}{|c|}{\textbf{Classifier}} & \textbf{1-gram} & \textbf{2-grams} & \textbf{3-grams} & \textbf{4-grams} \\
    \hline
    \textbf{NB} & 0.756 & 0.778 & 0.784 & \textbf{0.785} \\
    \hline
    \textbf{LR} & 0.821 & 0.835 & \textbf{0.836} & \textbf{0.836} \\
    \hline
    \textbf{SVM} & 0.823 & 0.835 & 0.836 & \textbf{0.837} \\
    \hline
    \textbf{RF} & 0.838 & \textbf{0.84} & 0.838 & 0.838 \\
    \hline
    \textbf{XGB} & 0.844 & 0.845 & \textbf{0.846} & \textbf{0.846} \\
    \hline
    \textbf{LGBM} & \textbf{0.845} & \textbf{0.845} & \textbf{0.845} & \textbf{0.845} \\
    \hline
    \end{tabular}%
  \label{tab:ngrams_clf}%
\end{table}%

\begin{figure}[t]
    \centering
    {%
    \setlength{\fboxsep}{0pt}%
    \setlength{\fboxrule}{0.5pt}%
    \fbox{\includegraphics[width=8.3cm,keepaspectratio]{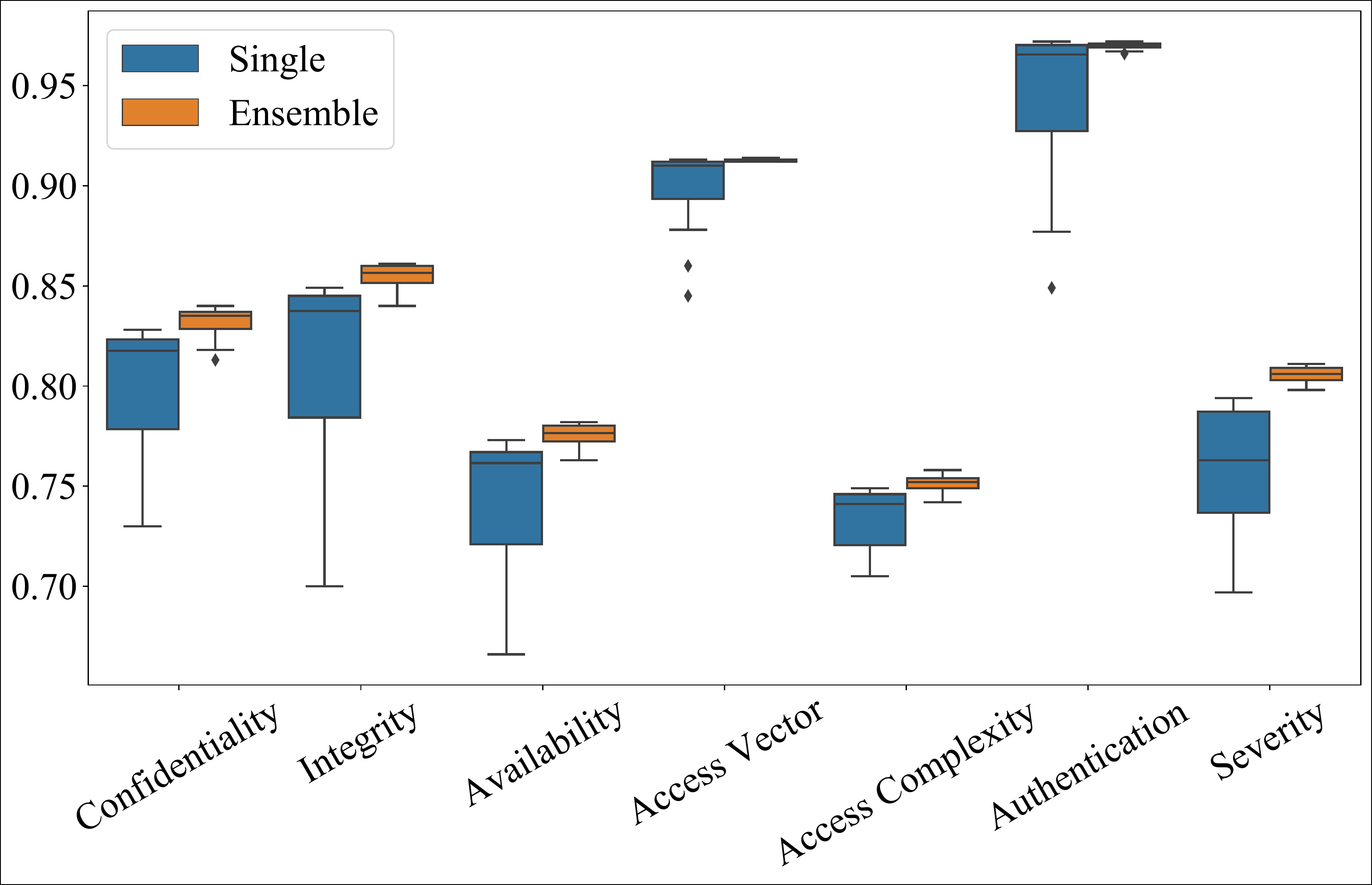}
    }}
    \caption{Average cross-validated \emph{Weighted F-Score}s comparison between ensemble and single models for each VC.}
    \label{fig:clf_types}
\end{figure}

Besides each output, we also examined the validated results across different types of classifiers (single vs. ensemble models) and NLP representations (n-grams and tf-idf vs. term frequency). Since the NLP representations mostly affect the classifiers, their validated results are grouped by six classifiers in Table~\ref{tab:feature_clf} and Table~\ref{tab:ngrams_clf}. The result shows that tf-idf did not outperform term frequency for five out of six classifiers. This result agrees with the existing work~\cite{spanos2018multi,spanos2017assessment}. It seemed that n-grams with n $>$ 1 improved the result. We used a right-tailed unpaired two-sample t-test to check the significance of such improvement of n-grams (n $>$ 1). The P-value was 0.169; that was larger than the confidence level of 0.05. Thus, we were unable to accept the improvement of n-grams over uni-gram. Furthermore, there was no performance improvement after increasing the number of n-grams. The above-reported three observations implied that the more complex NLP representations did not provide a statistically significant improvement over the simplest BoW (configuration one in Table~\ref{tab:model_configurations}). This argument helped explain why three out of seven optimal models in Table~\ref{tab:optimal_models} were BoW.

\begin{table}[b]
    \centering
    \caption{Average cross-validated \emph{Weighted F-scores} of uni-gram vs. n-grams (2 $\leq$ n $\leq$ 4) grouped by six classifiers.}
    \begin{tabular}{|l|c|}
        \hline
        \textbf{Vulnerability characteristic} & \textbf{P-value}       \\ \hline
        \textbf{Confidentiality} & $3.261 \times 10^{-5}$ \\ \hline
        \textbf{Integrity} & $9.719 \times 10^{-5}$  \\ \hline
        \textbf{Availability} & $3.855 \times 10^{-5}$  \\ \hline
        \textbf{Access Vector} & $2.320 \times 10^{-3}$  \\ \hline
        \textbf{Access Complexity} & $1.430 \times 10^{-5}$  \\ \hline
        \textbf{Authentication} & $1.670 \times 10^{-3}$  \\ \hline
        \textbf{Severity} & $1.060 \times 10^{-7}$  \\ \hline
    \end{tabular}
    \label{tab:clf_test}
\end{table}

Along with the NLP representations, we also investigated the performance difference between single (NB, LR, and SVM) and ensemble (RF, XGB, and LGBM) models. The average \emph{Weighted F-Scores} grouped by VCs for single and ensemble models are illustrated in Fig.~\ref{fig:clf_types}. Ensemble models seemed to consistently demonstrate the superior performance compared to single counterparts. It was also observed that the ensemble methods produced mostly consistent results (i.e., small variance) for Access Vector and Authentication properties. We performed the right-tailed unpaired two-sample t-tests to check the significance of the better performance of ensemble over single models. Table~\ref{tab:clf_test} reports the P-values of the results from the hypothesis testing. The hypothesis testing confirmed that the superiority of the ensemble methods was significant since all P-values are smaller than the confidence level of 0.05. The validated results in Table~\ref{tab:optimal_models} also affirmed that six out of seven optimal classifiers were ensemble (i.e., LGBM and XGB). It is noted that the XGB model usually takes more time to train than the LGBM model, especially for tf-idf representation. Our findings suggest that LGBM, XGB and BoW should be considered as baseline classifiers and NLP representations for future vulnerability-related research.

\begin{tcolorbox}
\textbf{The summary answer to RQ2}: LGBM and BoW are the most frequent optimal classifiers and NLP representations. Overall, the more complex NLP representations such as n-grams, tf-idf do not provide a statistically significant performance improvement than BoW. The ensemble models perform statistically better than single ones. It is recommended that the ensemble classifiers (e.g., XGB and LGBM) and BoW should be used as baseline models for vulnerability analytics.
\end{tcolorbox}

\subsection{RQ3: How effective is our character-word model to perform automated vulnerability assessment with concept drift?}
\label{subsec:rq3_results}

The OoV terms presented in RQ1 actually directly have an impact on the word-only models. Such missing features can make the model unable to produce reliable results. Especially when no existing term is found (i.e., all features are zero), the model would have the same output regardless of the context. To answer RQ3, we first tried to identify such all-zero cases in the vulnerability descriptions from 2000 to 2018. For each year from 2000 to 2018, we split the dataset into (\emph{i}) training set (data from the previous year backward) for building the vocabulary, and (\emph{ii}) testing set (data from the current year to 2018) for checking the vocabulary existence. We found 64 cases from 2000 to 2018 in the testing data, in which all the features were missing (cf. Appendix~\ref{sec:Appendix}). We used the terms appearing at least 0.1\% in all descriptions. It should be noted that the number of all-zero cases may be reduced using a larger vocabulary with the trade-off for larger computational time. We also investigated the descriptions of these vulnerabilities and found several interesting patterns. The average length of these abnormal descriptions was only 7.98 words compared to 39.17 of all descriptions. It turned out that the information about the threats and sources of such SVs was limited. Most of them just included the assets and attack/vulnerability types. For example, the vulnerabilities with ID of CVE-2016-10001xx had nearly the same format ``Reflected XSS in WordPress plugin'' with the only differences were the name and version of the plugin. This format made the model hard to evaluate the impact of each vulnerability separately. Another issue was due to the specialized or abbreviated terms such as $/redirect?url= XSS, SEGV, CSRF$ without proper explanation. The above issues suggest that the vulnerability descriptions should be written with sufficient information to enhance the comprehensibility of SVs.

For RQ3, the solution to the issue of the word-only model using character level features is evaluated. We considered the non-stop-words with high frequency (i.e., appearing in more than 10\% of all descriptions) to generate the character features. Using the same 0.1\% value as RQ2 increased the dimensions more than 30 times, but the performance only changed within 0.02. According to Algorithm~\ref{algo:feature_aggregation}, the output minimum number of character n-grams was chosen to be two. We first tested the robustness of the character-only models by setting the maximum number of characters to only three. For each year y from 1999 to 2017, we used such character model to generate the characters from the data of the considering year \emph{y} backward. We then verified the existence of such features using the descriptions of the other part of data (i.e., from year \emph{y} + 1 towards 2018). Surprisingly, the model using only two-to-three-character n-grams could produce at least one non-zero feature for all the descriptions even using only training data in 1999 (i.e., the first year in our dataset based on the vulnerability identification). Such finding shows that our approach is stable to vulnerability data changes (\emph{concept drift}) in testing data from 2000 to 2018 even with the limited amount of data and without retraining.

\begin{figure}[t]
    \centering
    {%
    \setlength{\fboxsep}{0pt}%
    \setlength{\fboxrule}{0.5pt}%
    \fbox{\includegraphics[width=8.3cm,keepaspectratio]{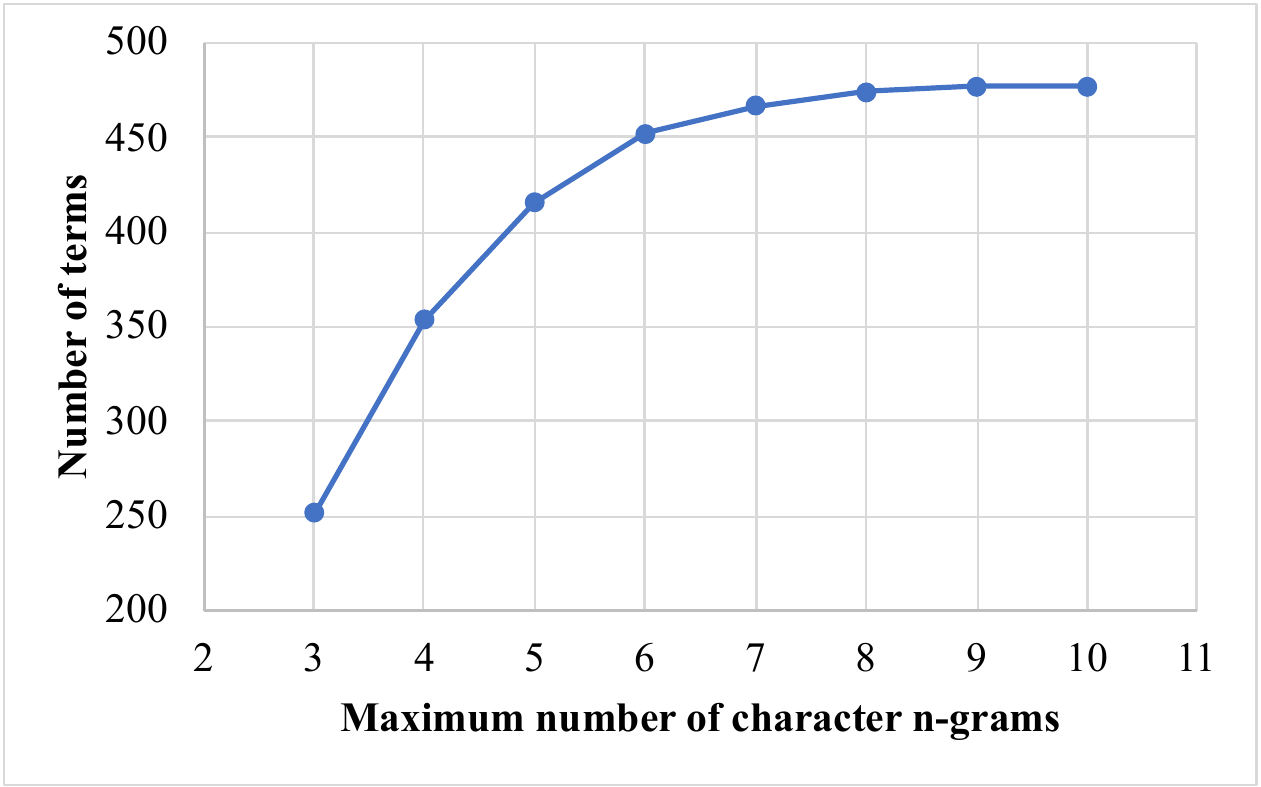}
    }}
    \caption{The relationship between the size of vocabulary and the maximum number of character n-grams.}
    \label{fig:char_ngrams}
\end{figure}

Next, to increase the generalizability of our approach, three to ten were considered for selecting the maximum number of character n-grams based on their corresponding vocabulary sizes (cf. Fig.~\ref{fig:char_ngrams}). Using the elbow method in cluster analysis, six was selected since vocabulary size did not increase dramatically after this point. The selected minimum and maximum values of character n-grams turned out to match the minimum and average word lengths of all NVD descriptions in our dataset, respectively.

\begin{table*}[t]
  \centering
  \caption{Performance (\emph{Accuracy}, \emph{Macro F-Score}, \emph{Weighted F-Score}) of our character-word, word-only and character-only models.}
    \begin{tabular}{|l|c|c|c|c|c|c|c|c|c|}
    \hline
    \multicolumn{1}{|c|}{\multirow{2}{*}{\textbf{Vulnerability characteristic}}} & \multicolumn{3}{c|}{\textbf{Our optimal model (CWM)}} & \multicolumn{3}{c|}{\textbf{Word-only model (WoM)}} & \multicolumn{3}{c|}{\textbf{Character-only model (CoM)}} \\
    \cline{2-10}
     & \textbf{\emph{Accuracy}} & \makecell{\textbf{\emph{Macro}}\\ \textbf{\emph{F-Score}}} & \makecell{\textbf{\emph{Weighted}}\\ \textbf{\emph{F-Score}}} & \textbf{\emph{Accuracy}} & \makecell{\textbf{\emph{Macro}}\\ \textbf{\emph{F-Score}}} & \makecell{\textbf{\emph{Weighted}}\\ \textbf{\emph{F-Score}}} & \textbf{\emph{Accuracy}} & \makecell{\textbf{\emph{Macro}}\\ \textbf{\emph{F-Score}}} & \makecell{\textbf{\emph{Weighted}}\\ \textbf{\emph{F-Score}}} \\
    \hline
    \textbf{Confidentiality} & \textbf{0.727} & \textbf{0.717} & \textbf{0.728} & 0.722 & 0.708 & 0.723 & 0.694 & 0.683 & 0.698 \\
    \hline
    \textbf{Integrity} & 0.763 & \textbf{0.749} & 0.764 & 0.763 & 0.744 & 0.764 & 0.731 & 0.718 & 0.734 \\
    \hline
    \textbf{Availability} & \textbf{0.712} & \textbf{0.711} & \textbf{0.711} & 0.700 & 0.696 & 0.702 & 0.660 & 0.657 & 0.660 \\
    \hline
    \textbf{Access Vector} & \textbf{0.914} & \textbf{0.540} & \textbf{0.901} & 0.904 & 0.533 & 0.894 & 0.910 & 0.538 & 0.899 \\
    \hline
    \textbf{Access Complexity} & 0.703 & 0.468 & 0.673 & \textbf{0.718} & \textbf{0.476} & \textbf{0.691} & 0.700 & 0.457 & 0.668 \\
    \hline
    \textbf{Authentication} & \textbf{0.875} & \textbf{0.442} & \textbf{0.844} & 0.864 & 0.425 & 0.832 & 0.866 & 0.441 & 0.840 \\
    \hline
    \textbf{Severity} & 0.668 & \textbf{0.575} & 0.663 & \textbf{0.686} & 0.569 & \textbf{0.675} & 0.661 & 0.549 & 0.652 \\
    \hline
    \end{tabular}%
  \label{tab:baseline_comparison}%
\end{table*}%

We then used the feature aggregation algorithm (cf. section~\ref{subsec:feature_aggregation}) to create the aggregated features from the character n-grams (2 $\leq$ n $\leq$ 6) and word n-grams to build the final model set and compared it with two baselines: Word-only Model (WoM) and Character-only Model (CoM). It should be noted that WoM is the model in which \emph{concept drift} is not handled. Unfortunately, a direct comparison with the existing WoM [6] was not possible since they used older NVD dataset and they did not produce their source code for reproduction. However, we tried to set up the experiments based on the guidelines and results in the previous paper.

To be more specific, we used BoW predictors and random forest (the best of their three models used) with the following hyperparameters: the number of trees was 100 and the number of features for splitting was 40. For CoM, we used the same optimal classifiers of each VC. The comparison results are given in Table~\ref{tab:baseline_comparison}. CWM performed slightly better than the WoM for four out of seven VCs regarding all evaluation metrics. Also, 4.98\% features of CWM were non-zero, which was nearly five-time denser than 1.03\% of WoM. Also, CoM was the worst model among the three, which had been expected since it contained the least information (smallest number of features). Although CWM does not significantly outperform WoM, its main advantage is to effectively handle the OoV terms (\emph{concept drift}), except new terms without any matching parts. We hope that our solution to \emph{concept drift} will be integrated into the practitioner's existing framework and future research work to perform more robust SVs analytics.

\begin{tcolorbox}
\textbf{The summary answer to RQ3}: The WoM does not handle the new cases well, especially those with all zero-value features. Without retraining, the tri-gram character features can still handle the OoV words effectively with no all-zero features for all testing data from 2000 to 2018. Our CWM performs comparably well with the existing WoM and provides nearly five-time richer information. Hence, our CWM is better for automated vulnerability assessment with \emph{concept drift}.
\end{tcolorbox}

\subsection{RQ4: To what extent can low-dimensional model retain the original performance?}
\label{subsec:rq4_results}

The n-gram NLP models usually have an issue with the high-dimensional and sparse feature vectors~\cite{kao2007natural}. The large feature sizes of our CWMs in Table~\ref{tab:baseline_comparison} were 1649 for Confidentiality, Availability and Access Complexity; 4154 for Integrity and Access Vector; 3062 for Authentication; and 5104 for Severity. To address such challenge in RQ4, we investigated the dimensionality reduction method (i.e., Latent Semantic Analysis (LSA)~\cite{deerwester1990indexing}) and recent sub-word embeddings (e.g., fastText~\cite{bojanowski2017enriching,joulin2016bag}) for vulnerability classification. fastText is an extension of Word2Vec~\cite{mikolov2013distributed} word embeddings, in which the character-level features are also considered. fastText is different to traditional n-grams in the sense that it determines the meaning of a word/subword based on the surrounding context. Here, we computed the sentence representation as an average fastText embedding of its constituent words and characters. We implemented fastText using \emph{Gensim}~\cite{rehurek2010software} library in Python.

\begin{table}[t]
  \centering
  \caption{\emph{Weighted F-Scores} of our original CWM (green-colored baseline), 300-dimension Latent Semantic Analysis (LSA-300), fastText trained on vulnerability description (fastText-300) and fastText trained on English Wikipedia pages (fastText-300W).}
    \begin{tabular}{|l|c|c|c|c|}
    \hline
    \makecell[l]{\textbf{Vulnerability}\\ \textbf{characteristic}} & \makecell{\textbf{Our}\\ \textbf{CWM}} & \makecell{\textbf{LSA-}\\ \textbf{300}} & \makecell{\textbf{fastText-}\\ \textbf{300}} & \makecell{\textbf{fastText-}\\ \textbf{300W}} \\
    \hline
    \textbf{Confidentiality} & {\color[HTML]{008000}\textbf{0.728}} & 0.656 & \textbf{0.679} & 0.648 \\
    \hline
    \textbf{Integrity} & {\color[HTML]{008000}\textbf{0.764}} & 0.695 & \textbf{0.719} & 0.672 \\
    \hline
    \textbf{Availability} & {\color[HTML]{008000}\textbf{0.711}} & 0.656 & \textbf{0.687} & 0.669 \\
    \hline
    \textbf{Access Vector} & {\color[HTML]{008000}\textbf{0.901}} & 0.892 & \textbf{0.893} & 0.866 \\
    \hline
    \textbf{Access Complexity} & {\color[HTML]{008000}\textbf{0.673}} & 0.611 & \textbf{0.679} & 0.678 \\
    \hline
    \textbf{Authentication} & {\color[HTML]{008000}\textbf{0.844}} & \textbf{0.842} & 0.815 & 0.765 \\
    \hline
    \textbf{Severity} & {\color[HTML]{008000}\textbf{0.663}} & \textbf{0.656} & 0.654 & 0.635 \\
    \hline
    \end{tabular}%
  \label{tab:dimension_reduction}%
\end{table}%

For LSA, using the elbow method and total explained variances of the principal components, we selected 300 for the dimensions and called it LSA-300. Table~\ref{tab:dimension_reduction} shows that LSA-300 retained from 90\% to 99\% performance of the original model, but used only 300 dimensions (6-18\% original model sizes). More remarkably, with the same 300 dimensions, the fastText model trained on the vulnerability descriptions was on average better than LSA-300 (97\% vs. 94.5\%). fastText model even slightly outperformed our original CWM for Access Complexity. Moreover, for all seven cases, the fastText model using vulnerability knowledge (fastText-300) had higher \emph{Weighted F-Scores} than that trained on English Wikipedia pages (fastText-300W)~\cite{fasttest_pretrained}. The result implied that vulnerability descriptions contain specific terms that do not frequently appear in the general domains. The domain relevance turns out to be not only applicable to word embeddings~\cite{han2017learning}, but also to character/sub-word embeddings for vulnerability analysis and assessment. Overall, our findings show that LSA and fastText are capable of building efficient models without too much performance trade-off.

\begin{tcolorbox}
\textbf{The summary answer to RQ4}: The LSA model with 300 dimensions (6-18\% of the original size) retains from 90\% up to 99\% performance of the original model. With the same dimensions, the model with fastText sub-word embeddings provide even more promising results. The fastText model with the vulnerability knowledge outperforms that trained on the general context (e.g., Wikipedia). LSA and fastText can help build efficient models for vulnerability assessment.
\end{tcolorbox}

\section{Threats to Validity}
\label{sec:discussion}

\noindent \textbf{Internal validity}. We used well-known tools such as \emph{scikit-learn} [26] and \emph{nltk} [27] libraries for ML and NLP. Our optimal models may not guarantee the highest performance for every SV since there are infinite values of hyperparameters to tune. However, even when the optimal values change, a time-based cross-validation method should still be preferred since we have considered the general trend of all SVs. Our model may not provide the state-of-the-art results, but at least it gives the baseline performance for handling the \emph{concept drift} of SVs.

\noindent \textbf{External validity}. Our work used NVD -- one of the most comprehensive public repositories of SVs. The size of our dataset is more than 100,000 with the latest vulnerabilities in 2018. Our character-word model has also been demonstrated to consistently handle the OoV words well even with very limited data for all years in the dataset. It is recognized that the model may not work for extreme rare terms in which no existing parts can be found. However, our model is totally re-trainable to deal with such cases or incorporate more sources of SVs' descriptions.

\noindent \textbf{Conclusion validity}. We mitigated the randomness of the results by taking the average value of five-fold cross-validation. The performance comparison of different types of classifiers and NLP representations was also confirmed using a statistical hypothesis test with P-values that were much lower than the confidence level of 5\%.

\section{RELATED WORK}
\label{sec:related_work}

\subsection{Software Vulnerability Analytics}

It is important to patch the critical-first vulnerabilities~\cite{hou2017security}. Thus, besides CVSS, there have been many effective evaluation schemes for SVs~\cite{liu2011vrss,spanos2013wivss,sharma2018improved}. Recently, there is a detailed Bayesian analysis of various vulnerability scoring systems~\cite{johnson2016can}, which highlights the good overall performance of CVSS. Therefore, we used the well-known CVSS as the ground truth for our approach. We assert that our approach can be generalizable to other vulnerability rating systems following the same scheme of multi-class classification.

Regarding the predictive analytics of SVs, Bozorgi~\cite{bozorgi2010beyond} pioneered the use of ML models for SVs analysis. Their paper used an SVM model and various features (e.g., NVD description, CVSS, published and modified dates) to estimate the likelihood of exploitation and \emph{time-to-exploit} of SVs. Another piece of work analyzed the VCs and trends of SVs by incorporating different vulnerability information from multiple vulnerability repositories~\cite{murtaza2016mining,almukaynizi2019patch}, security advisories~\cite{edkrantz2015predicting,huang2013novel}, darkweb/deepnet~\cite{almukaynizi2019patch,nunes2016darknet} and social network (Twitter)~\cite{sabottke2015vulnerability}. These efforts assumed that all VCs have been available at the time of analysis. However, our work relaxes this assumption by using only the vulnerability description – one of the first information about new SVs. Our model can be used for both new and old SVs.

Actually, the descriptions are also utilized for vulnerability analysis and prediction. Yamamoto~\cite{yamamoto2015text} used Linear Discriminant Analysis, Na\"ive Bayes and Latent Semantic Indexing combined with annual effect estimation to determine the VCs of more than 60,000 SVs in NVD. The annual effect focused on the recent SVs, but still could not explicitly handle the OoV terms in the descriptions. Spanos~\cite{spanos2018multi} worked on the same task using a multi-target framework with Decision Tree, Random Forest and Gradient Boosting Tree. Our approach also contains the word-only model, but we select the optimal models using our time-based cross-validation to better address the \emph{concept drift} issue. The vulnerability descriptions were also used to evaluate the vulnerability severity~\cite{spanos2017assessment}, associate the frequent terms with each VC~\cite{toloudis2016associating}, determine the type of each SV using topic modeling~\cite{neuhaus2010security} and show vulnerability trends~\cite{murtaza2016mining}. Recently, Zhuobing [8] have applied deep learning to predict vulnerability severity. The existing literature has demonstrated the usefulness of description for vulnerability analysis and assessment, but has not mentioned how to overcome its \emph{concept drift} challenge. Our work is the first of its kind to provide a robust treatment for SVs' \emph{concept drift}.

\subsection{Temporal Modeling of Software Vulnerabilities}

Regarding the temporal relationship of SVs, Roumani~\cite{roumani2015time} proposed a time-series approach using autoregressive integrated moving average and exponential smoothing methods to predict the number of vulnerabilities in the future. Another time-series work~\cite{tang2016exploiting} was described to model the trend in disclosing SVs. A group of researchers led by Tsokos published a series of work~\cite{rajasooriya2017cyber,kaluarachchi2017non,pokhrel2017cybersecurity} on stochastic models such as Hidden Markov Models, Artificial Neural Network, and Support Vector Machine to estimate the occurrence and exploitability of vulnerabilities. The focus of the above studies was on the determination of the occurrence of SVs over time. In contrast, our work aims to handle the temporal relationship to build more robust predictive models for SVs assessment.

\section{CONCLUSIONS AND FUTURE WORK}
\label{sec:conclusions}

We observe that the existing works suffer from \emph{concept drift} in the vulnerability descriptions that affect both the traditional model selection and prediction steps of SVs assessment. We assert that \emph{concept drift} can degrade the robustness of existing predictive models. We show that the time-based cross-validation should be used for vulnerability analysis to better capture the temporal relationship of SVs. Then, our main contribution is the Character-Word Models (CWMs) to improve the robustness of automated SVs assessment with \emph{concept drift}. CWMs have been demonstrated to handle \emph{concept drift} of SVs effectively for all the testing data from 2000 to 2018 in NVD even in the case of data scarcity. Our approach has also performed comparably well with the existing word-only models. Our CWMs are also much less sparse and thus less prone to overfitting. We have also found that Latent Semantic Analysis and sub-word embeddings like fastText help build compact and efficient CWM models (up to 94\% reduction in dimension) with the ability to retain at least 90\% of the performance for all VCs. Besides the good performance, the implications on the use of different models are also given to support practitioners and researchers with vulnerability analytics. Hopefully, this work can open up various research avenues to develop more sophisticated \emph{concept-drift}-aware models in SVs and related areas.

In the future, we plan to investigate the performance of deep learning models to embed the dependency of both character and word features in low-dimensional space for vulnerability prediction. Alongside \emph{concept drift}, handling imbalanced data is also a concern for future SVs research.

\appendix

\section{Appendix}
\label{sec:Appendix}
\tiny
\noindent 64 vulnerabilities (CVD-ID) with all-zero features of word-only model (V.C) from 2000 to 2018:
CVE-2013-6647, CVE-2015-1000004, CVE-2016-1000113, CVE-2016-1000114, CVE-2016-1000117, CVE-2016-1000118, CVE-2016-1000126, CVE-2016-1000127, CVE-2016-1000128, CVE-2016-1000129, CVE-2016-1000130, CVE-2016-1000131, CVE-2016-1000132, CVE-2016-1000133, CVE-2016-1000134, CVE-2016-1000135, CVE-2016-1000136, CVE-2016-1000137, CVE-2016-1000138, CVE-2016-1000139, CVE-2016-1000140, CVE-2016-1000141, CVE-2016-1000142, CVE-2016-1000143, CVE-2016-1000144, CVE-2016-1000145, CVE-2016-1000146, CVE-2016-1000147, CVE-2016-1000148, CVE-2016-1000149, CVE-2016-1000150, CVE-2016-1000151, CVE-2016-1000152, CVE-2016-1000153, CVE-2016-1000154, CVE-2016-1000155, CVE-2016-1000217, CVE-2017-10798, CVE-2017-10801, CVE-2017-14036, CVE-2017-14536, CVE-2017-15808, CVE-2017-16760, CVE-2017-16785, CVE-2017-17499, CVE-2017-17703, CVE-2017-17774, CVE-2017-6102, CVE-2017-7276, CVE-2017-8783, CVE-2018-10030, CVE-2018-10031, CVE-2018-10382, CVE-2018-11120, CVE-2018-11405, CVE-2018-12501, CVE-2018-13997, CVE-2018-14382, CVE-2018-5285, CVE-2018-5361, CVE-2018-6467, CVE-2018-6834, CVE-2018-8817, CVE-2018-9130


\IEEEtriggeratref{49}

\bibliographystyle{IEEEtran}
\bibliography{reference}

\end{document}